\documentstyle[12pt]{article}
\begin{document}
\baselineskip 8mm
\title{Registering Seconds with a {\it Conic Clock}}
\author{
Ricardo L\'{o}pez-Ruiz$^{\ddag}$ and Amalio F. Pacheco$^{\dag}$,\\
 \small $^{\ddag}$ Departament of Computer Sciences and BIFI, \\
 \small $^{\dag}$ Departament of Theoretical Physics and BIFI, \\
 \small Faculty of Sciences - University of Zaragoza, \\
 \small 50009 - Zaragoza (Spain).
 }
 \date{ }

\maketitle
\begin{center} {\bf Abstract} \end{center}
The feasibility of registering seconds using the frictionless 
motion of a point-like particle that
slides under gravity on an inverted conical surface is studied.
Depending on the integer part of the relation between the angular 
and  radial frequencies of the particle trajectory,
only an angular interval for the cone is 
available for this purpose. 
For each one of these possible angles, there exists 
a unique trajectory that has the capability of registering seconds.
The method to obtain the geometrical properties of these trajectories 
and the necessary initial conditions to reach them are then established.

{\small {\bf Electronic mail:} $^{\ddag}$ rilopez@posta.unizar.es ;
$^{\dag}$ amalio@posta.unizar.es}

\newpage

\section{Introduction}
\label{sec:intro}

A clock, by definition, is an instrument that
 requires some kind of periodic process that registers 
the passing of the time. This periodicity can be typically obtained using 
an oscillating system such as a pendulum or a balance wheel. It also needs  
a trigger mechanism connecting the oscillating structure with a source of energy, 
such as a weight or spring, in order to compensate for the dissipation \cite{andronov}.
However, in a first approach, one could think of different ideal configurations in which
the bounded frictionless dynamics of a particle is used for registering 
time. The simplest example of this type is a particle bouncing on the ground under 
the sole effect of gravity. In this case, the relation 
between the height, $h$, from where the ball is released and the the time of a 
complete oscillation, $T$, is $h=gT^2/8$, with $g$ being gravity. 
Hence, if seconds is what we want to register, 
advice of at least $1.2$ metres height is necessary. 
In this case the oscillations are performed in a parallel direction to gravity.
On the contrary, the usual simple pendulum oscillates in a perpendicular direction
to the gravitational field due to the effect of the motionless point where the pendulum is fixed.
The relation, $l=gT^2/(4\pi^2)$, between the length of the pendulum, $l$, 
and the small-amplitude oscillation period, $T$, indicates that a $0.25$ metre pendulum-length
is required for registering seconds. In both cases the particle always bounces
on the same point. Then an additional mechanism should be inserted in these ideal clocks in order
to have, for instance, a moving light signal giving us the visual impression 
of the passing of time. 

A particle sliding on an inverted cone is an alternative system
that combines both oscillations, in the parallel and 
in the perpendicular directions to gravity, and 
that can put in evidence the passing of time without additional mechanisms. 
Think, for instance, of a phosphorescent circular limb at
the top of the cone where the recurrent trajectory of the particle is bouncing.
One second elapses between consecutive bounces, and thus
the reading of time is manifest. 
This hypothetical machine will be called
a {\it conic clock}. It can be inferred from the other cases,
and by dimensional analysis,
that the height, $H$, of the cone trunk where the particle motion develops, 
follows a similar law, $H=cgT^2$, 
with $c$ a constant of order $1$ depending on the opening angle of the cone.

In Section 2 we will recall the non-dimensional form of the evolution equations
of a particle sliding on a conical surface. Section 3 is devoted
to explaining the method to obtain the trajectories with a recurrence time of seconds.
Finally, our conclusions are stated in Section 4.

\section{Dynamics of the particle}

The detailed analysis of the frictionless motion of a particle
on a conical surface was developed in Ref. \cite{lopezruiz}.
Here we sketch the way to obtain the non-dimensional equations
of its dynamical evolution. The dynamics of the particle
inside the cone can be expressed in the generalized coordinates
$(r,\theta)$, where $r$ is the coordinate in the direction 
of cone generatrix and $\theta$ is
the angular variable around the vertical axis. 
The Lagrangian function ${\cal L}$ for this system in these
particular coordinates appears as
\begin{equation}
{\cal L} = {1\over 2}\; m(\dot{r}^2 +
r^2\dot{\theta}^2\sin^2\phi_0) - mgr\cos\phi_0, 
\label{eq:L}
\end{equation}
and the Hamiltonian, $\cal H$, is
\begin{equation}
{\cal H} = \frac{p_r^2}{2m} +
\frac{p_{\theta}^2}{2mr^2\sin^2\phi_0} + mgr\cos\phi_0.
\label{eq:H}
\end{equation}
The equations of motion of the particle are determined by the
Lagrange equations,
$\frac{d}{dt}\left(\frac{\partial\cal L}{\partial\dot{q}_i}\right) -
\frac{\partial\cal L}{\partial q_i} = 0 \label{eq:Lagrange}$,
with $q_i=(r,\theta)$, $i=1,2$.
The first of these equations for $q_1=r$ gives us the evolution of the
particle in the radial direction. It yields
\begin{equation}
\ddot{r} - r\dot{\theta}^2\sin^2\phi_0 + g\cos\phi_0 = 0.
\label{eq:radial1}
\end{equation}
And the second for $q_2=\theta$ 
puts in evidence an invariant of motion because
$\cal L$ is independent of the angular variable $\theta$:
\begin{equation}
mr^2\dot{\theta}\sin^2\phi_0 = cte. = L_z.
\label{eq:angular}
\end{equation}
This dynamical constant, $L_z$, is the vertical component of the
angular momentum.\newline If we substitute the value of
$\dot{\theta}=L_z/(mr^2\sin^2\phi_0)$ in Eq. (\ref{eq:radial1}),
the radial evolution is uncoupled from its angular dependence,
which remains present only through the constant $L_z$,
\begin{equation}
\ddot{r} - \left(\frac{L_z}{m\sin\phi_0}\right)^2\frac{1}{r^3} +
g\cos\phi_0 = 0.
 \label{eq:radial}
\end{equation}
This last equation corresponds to an integrable nonlinear
oscillator in the radial direction. After integrating this motion,
the angular part of the dynamics is obtained through Eq.
(\ref{eq:angular}). Note that the trajectory
is independent of the mass of the particle.

\subsection{Radial and angular frequency relation}
Given a $L_z$ there exists a unique circular orbit with this vertical component
of the angular momentum. The energy of this orbit is
$E_0=\frac{3}{2}\,mgr_0\cos\phi_0$, with $r_0$ its radial dimension,
\begin{equation}
r_0 = \left[\frac{L_z^2}{m^2g\sin^2\phi_0\cos\phi_0}\right]^{1\over 3},
\label{eq:r_0}
\end{equation}
and its angular frequency, $\omega_0$, given by
\begin{equation}
\omega_0^2 = \frac{g\cos\phi_0}{r_0\sin^2\phi_0}.\label{eq:w_0}
\end{equation}
A radial perturbation of this trajectory, that is, 
an increase in the energy, $E$, of the particle leaving $L_z$ constant, 
provokes an additional radial oscillation superimposed on the angular oscillation.
If $E$ is only slightly bigger than $E_0$, then the radial frequency is
\begin{equation}
\omega_{r,E_0}^2 = \frac{3g\cos\phi_0}{r_0} ,
\label{eq:wr0}
\end{equation}
and the relation between these two frequencies is
\begin{equation}
\frac{w_{r,E_0}}{w_0} = \sqrt{3} \sin{\phi_0}. \label{eq:rel-w}
\end{equation}
The trajectory of the perturbed circular orbit in the plane
$(\theta, r)$ covers a longer $\theta$-angular 
distance when its radial coordinate is under the value $r=r_0$ 
than when it is over $r_0$. 
When the energy increases, in the limit $E\rightarrow\infty$,
the dynamics for $r>r_0$ is
projected onto a tight peak in the $(\theta,r)$ plane. Thus most of
the part of the $\theta$-coordinate is covered when the particle
is under the circular orbit $r=r_0$ although the system spends
its time essentially over the circular orbit. In this limit 
the ratio of frequencies is:
\begin{equation}
\frac{\omega_{r,E=\infty}}{\omega_{\theta}} \;\; = \;\; 2\sin\phi_0.
\end{equation}

The general behaviour of this ratio, for an arbitrary value of $E$,
 was shown to be \cite{lopezruiz}:
\begin{equation}
 \label{eq:univ-rel}
\frac{\omega_r}{\omega_{\theta}} \; = \; k\sin\phi_0
\;\;\;\hbox{with $k$ in the range}\;\; \left\{\begin{array}{c}
 \sqrt{3} < k < 2 \\
 \;\;\updownarrow \mbox{ } \;\;\;\updownarrow \mbox{ } \;\;\updownarrow \\
\;\;\; E_0 < E < \infty  \end{array}\right. .
\end{equation}

\subsection{The non-dimensional equations}
The non-dimensional equations are obtained by rescaling
the radial, angular and time variables. 
If we substitute the value of the vertical component of the angular momentum,
$L_z^2$, given by Eq. (\ref{eq:r_0}) and perform the change of
variables: $\tilde{r}=r/r_0$, $\tilde{\theta}=\theta\sin\phi_0$
and $\tilde{t}=\frac{\omega_{r,E_0}}{\sqrt{3}}t$, the
radial and angular equations are reduced to
\begin{eqnarray}
\ddot{\tilde{r}} + (1 - \tilde{r}^{-3}) & = & 0, \label{eq1}\\
 \dot{\tilde{\theta}} - \tilde{r}^{-2} & = & 0. \label{eq2}
\end{eqnarray}
In this adimensional form, the equations apparently lose every 
characteristic of the system. Every possible trajectory on
the conical surface is projected into a solution of these equations
and, in that sense, we say that they are universal. They contain
all the information about the dynamical behaviour of the
particle.\newline In particular, $\tilde{r}= 1$ is the singularity
representing the circular orbit and any other orbit runs between
the extreme values, $\tilde{r}_{min}$ and $\tilde{r}_{max}$, which
satisfy: $0<\tilde{r}_{min}<1$ and $1<\tilde{r}_{max}<\infty$.
These extreme points verify the relation
\begin{equation}
\frac{2\,\tilde{r}_{min}^2\tilde{r}_{max}^2}{\tilde{r}_{min}+\tilde{r}_{max}}
= 1. \label{eq:traject-rel1}
\end{equation}

\section{Trajectories clocking seconds}

An important result in Hamiltonian dynamics is the {\it Poincar\'e recurrence theorem}.
It states that a bounded Hamiltonian system returns systematically to an arbitrarily small
neighborhood of its initial condition \cite{arnold,goldstein}. Apparently this microscopic reversibility
would imply the same behaviour in the macroscopic scale, and this is the case,
but this theorem does not give any
information about the time that the system takes to return close to its initial state.
As this recurrence time is extremely long in most of the real systems, 
the usual irreversible macroscopic behaviour is observed. 
In spite of the lack of a general rule to calculate the recurrence time, the success of
statistical mechanics has consisted in converting the ignorance of the recurrence times
of a system in equilibrium into a probabilistic value: that correspondent to the distribution 
or the ensemble associated to the Hamiltonian physical situation under study.

Our goal now consists in calculating exactly in this particular system
the motions of the particle with a recurrence time of seconds, $T=1$, and 
advancing a $2\pi\delta$ clockwise polar angle in each 
new bouncing on the top of the conic clock.
The first step is to set up the interval of angles in the opening of the cone,
$[\phi_{min},\phi_{max}]$, that can handle a trajectory verifying these conditions. 
Thus, the frequency relation must verify:
\begin{equation}
\frac{\omega_{\theta}}{\omega_r} \;\; = \;\; n + \delta,
\end{equation}
where $n$ is the number of complete angular rounds for each radial oscillation
and $\delta$ is the $2\pi$-clockwise ratio shift between two successive particle 
marks on the top surface of the conic clock. 
From Eq. (\ref{eq:univ-rel}) we find that the extreme angles are given by the relation
\begin{equation}
k\sin\phi_0=1/(n+\delta)
\label{eq:k1} 
\end{equation}
when $k=\sqrt{3}$ and $k=2$. The result is 
\begin{eqnarray}
\sin\phi_{max} & = & \frac{1}{\sqrt{3}(n+\delta)} ,\\
\sin\phi_{min} & = & \frac{1}{2(n+\delta)}. 
\end{eqnarray}
Thus, we see now that there exists a unique orbit for each 
$\phi_0\in [\phi_{min},\phi_{max}]$, when $T$, $n$ and $\delta$ are fixed.
For this $\phi_0$ there is a $k_0$ satisfying the relation (\ref{eq:k1}).
Because $k$ is a monotone increasing function
of the non-dimensional energy parameter $\tilde{E}=\frac{3E}{2E_0}$,
there exists a concrete value of $\tilde{E}_0$ which determines 
the trajectory in the non-dimensional 
Eqs. (\ref{eq1}-\ref{eq2}) or, equivalently, by undoing the change of variables,
the real motion in the Eqs. (\ref{eq:angular}-\ref{eq:radial}) .  
The function $k(\tilde{E})$ can be calculated by computational methods as follows. 

Firstly, we observe that the energy conservation condition, 
in the non-dimensional coordinates, becomes
\begin{equation}
\label{eq:radial-e}
\frac{\dot{\tilde{r}}^2}{2} + \tilde{V}(\tilde{r}) = \tilde{E},
\end{equation}
with $\tilde{V}(\tilde{r}) = \tilde{r}+\frac{1}{2\tilde{r}^2}$.
Hence, in this representation, the
dynamics settles in the circular orbit when $\tilde{E}=3/2$, and, small or
large oscillations are obtained when the value of the normalized
energy $\tilde{E}$ runs on the interval $\frac{3}{2}<\tilde{E}<\infty$.
For an arbitrary energy $\tilde{E}$, the extreme values, 
$\tilde{r}_{min}$ and $\tilde{r}_{max}$, of the radial oscillation are
deduced from the equality
$\tilde{V}(\tilde{r}_{min}) = \tilde{V}(\tilde{r}_{max})= \tilde{E}$.
Then the functions
$\tilde{r}_{min}(\tilde{E})$ and $\tilde{r}_{max}(\tilde{E})$ are obtained.
The angle $\tilde{\theta}$ covered by the particle during a
radial semi-period is given by
\begin{equation}
\label{eq:theta}
\tilde{\theta}(\tilde{E}) =
\int^{\tilde{r}_{max}(\tilde{E})}_{\tilde{r}_{min}(\tilde{E})}
\frac{d\tilde{r}}{\tilde{r}\sqrt{2\tilde{E}\tilde{r}^2-2\tilde{r}^3-1}},
\end{equation}
and $k(\tilde{E})$ is
\begin{equation}
\label{eq:k}
k(\tilde{E}) = \frac{\pi}{\tilde{\theta}(\tilde{E})}.
\end{equation}
Let us recall that $\sqrt{3}<k(\tilde{E})<2$,
for any trajectory of arbitrary energy, $\frac{3}{2}<\tilde{E}<\infty$.
In order to find the computational solution of the equation
$k(\tilde{E})=k_0$, we can start an iteration 
process using the initial values: $k=\sqrt{3}$ and $\tilde{E}=3/2$. 
By increasing $\tilde{E}$ in a fixed 
small quantity $\Delta\tilde{E}$, we calculate for each step a new $k$ 
through Eqs. (\ref{eq:theta}-\ref{eq:k}). When the condition $k>k_0$ is reached,
the iteration process stops and this value of $\tilde{E}$ is retained as
the solution for $\tilde{E}_0$. If more precision is required then it is sufficient to take a
smaller $\Delta\tilde{E}$.

Secondly, we proceed to find the period of the radial
oscillation, $\tilde{T}_{\tilde{r}}$, for the particular motion given by 
$\tilde{E}=\tilde{E}_0$. This is obtained by integrating the expression:
\begin{equation}
\tilde{T}_{\tilde{r}}(\tilde{E}_0) = 
2\int^{\tilde{r}_{max}(\tilde{E}_0)}_{\tilde{r}_{min}(\tilde{E}_0)} 
\frac{\tilde{r} d\tilde{r}}{\sqrt{2\tilde{E}_0\tilde{r}^2-2\tilde{r}^3-1}}.
\end{equation}

Now the change from the non-dimensional to the dimensional variables is performed.
Undoing the time change we obtain
\begin{equation}
r_0=g\cos\phi_0\left(\frac{T}{\tilde{T}_{\tilde{r}}(\tilde{E}_0)}\right)^2
\end{equation}
where $T=1$ second in our concrete case. From this value the radial and height lengths 
of the trajectory are obtained:
\begin{eqnarray}
r_{min} & = & \tilde{r}_{min}(\tilde{E}_0)\cdot r_0, \\
r_{max} & = & \tilde{r}_{max}(\tilde{E}_0)\cdot r_0, \\
H & = & (r_{max}-r_{min})\cdot \cos\phi_0,
\end{eqnarray}
with $H=H_{max}-H_{min}$ the height of the trunk of the cone 
where the dynamics develops.
The length, $L$, that the particle covers on the conical surface, 
in a radial semi-oscillation, is 
\begin{equation}
{L}(\tilde{E}_0) = 
r_0\cdot\int^{\tilde{r}_{max}(\tilde{E}_0)}_{\tilde{r}_{min}(\tilde{E}_0)} 
\sqrt{\frac{2\tilde{E}_0\tilde{r}^2-2\tilde{r}^3}{2\tilde{E}_0\tilde{r}^2-2\tilde{r}^3-1}}\;\;d\tilde{r}.
\end{equation}
If we take the $H$ and $L$ dependence on $r_0$, 
then, as we have advanced in the Introduction, it is found that
\begin{eqnarray}
H \sim cte_H(\phi_0)\cdot gT^2, \\
L \sim cte_L(\phi_0)\cdot gT^2,
\end{eqnarray} 
where $cte_H$ and $cte_L$ depend only on the opening angle of the cone.

Finally, the velocity of release at the highest point of the trajectory, 
$v_{\theta}$, can be obtained from the relation
\begin{equation}
v_{\theta}=\sqrt{\frac{2\gamma g H}{1-\gamma}}\;,
\end{equation}
with $\gamma=\left(\frac{r_{min}}{r_{max}}\right)^2$.

The results of this method, for  different values of $n$ and $\delta$,
are collected in Tables $1-3$. These calculations show that the fact
of clocking seconds, $T=1$, in the gravitational field imposes strong restrictions in 
the dimensions of the conic clock. If the mean value of $L$ is obtained in each table,
we observe that an approximate constant value of $c_L=L/(gT^2)$ appears 
in the three tables. In Table 1, $c_L\sim 0.109$, in Table 2, $c_L\sim 0.108$, and
in Table 3, $c_L\sim 0.124$. These non-dimensional constants are similar to those
commented on in Section I. In the case of the particle bouncing on the ground, $c_1=1/8=0.125$,
and that of the simple pendulum, $c_2=1/(4\pi^2)=0.025$.

\section{Conclusions}

It is known that the calculation of the recurrence time in 
a bounded Hamiltonian system is, in general, not an easy task.
In fact, the consequence of the difficulty for its 
computability gave rise to the well-known
polemics on the "recurrence paradox" at the end of the 
nineteenth century \cite{boltzmann,cohen}.
This discussion was concluded by Boltzmann who finally admitted that recurrences
are completely consistent with the statistical viewpoint and there is no contradiction
with the second law of thermodynamics. Thus recurrences can be interpreted as fluctuations,
which are almost certain to occur if a long enough time is waited. 

In this work, the orbits with a recurrence time of one second have been obtained
for the frictionless dynamics of a particle on an inverted cone. Depending on the
angular and radial frequency ratio, an interval of opening angles of the cone
is available for this purpose. For each one of these angles, only a trajectory
embodies the possibility of registering seconds. The method to calculate
these trajectories has been established.
We have also calculated the characteristic lengths of the conical 
surface for different values of the frequency ratio and for different opening angles.
It is worth remarking the strong restrictions that the gravitational field imposes on the possible 
dimensions of the hypothetical device registering seconds: if $L$ is the length
of the conic clock, $T$ is the time to be registered and $g$ is gravity, then 
the non-dimensional quantity $c_L=L/(gT^2)$ varies only a little for each interval of
possible angles. The value of this constant indicates that the dimensions of a conic clock
are of the same order as the dimensions of a clock-device based on the dynamics of
a particle bouncing on the ground or, more easily, based on the classical simple pendulum.
The advantage of the conic clock with respect to the other cases is that the particle does 
not bounces on the same point in each new recurrence. Therefore, it is not necessary 
to insert here an additional mechanism to obtain the visual impression of 
the passing of time.

\newpage

\begin{center} \bf{TABLES} \end{center}

\vspace{3cm}
\begin{table}[h]
\begin{center}
\begin{tabular}{|c|c|c|c|c|c|c|} \hline
$\phi_0$ & $\tilde{E}$ & $H_{min}$  & $H_{max}$ &   $H$   & $L$     & $v_{\theta}$  \\ \hline \hline
 $29.5$  &   $108.2$   &  $0.0006$  &  $0.928$  & $0.927$ & $1.066$ & $0.003$ \\ \hline
 $30$    &   $15.5$    &  $0.011$   &  $0.919$  & $0.909$ & $1.062$ & $0.049$ \\ \hline
 $31$    &   $6$       &  $0.044$   &  $0.894$  & $0.849$ & $1.054$ & $0.202$ \\ \hline
 $32$    &   $3.6$     &  $0.092$   &  $0.859$  & $0.767$ & $1.055$ & $0.419$ \\ \hline
 $33$    &   $2.5$     &  $0.165$   &  $0.805$  & $0.639$ & $1.065$ & $0.744$ \\ \hline
 $34$    &   $1.8$     &  $0.287$   &  $0.706$  & $0.419$ & $1.089$ & $1.276$ \\ \hline
 $34.5$  &   $1.55$    &  $0.412$   &  $0.594$  & $0.181$ & $1.107$ & $1.820$ \\ \hline
\end{tabular}
\caption{$T=1$, $n=1$ and $\delta=1/60$, then $\phi_{min}=29.46^o$ and 
$\phi_{max}=34.60^o$. The lengths are in metres and the velocity in metres/second.}
\label{tabla1}
\end{center}
\end{table}

\vspace{3cm}
\begin{table}[h]
\begin{center}
\begin{tabular}{|c|c|c|c|c|c|c|} \hline
$\phi_0$ & $\tilde{E}$ & $H_{min}$  & $H_{max}$ &   $H$   & $L$     & $v_{\theta}$  \\ \hline \hline
 $30.2$  &   $34.6$    &  $0.003$   &  $0.916$  & $0.912$ & $1.058$ & $0.015$ \\ \hline
 $30.6$  &   $14.2$    &  $0.012$   &  $0.908$  & $0.896$ & $1.056$ & $0.056$ \\ \hline
 $31$    &   $9.1$     &  $0.023$   &  $0.898$  & $0.874$ & $1.051$ & $0.108$ \\ \hline
 $32$    &   $4.8$     &  $0.061$   &  $0.870$  & $0.809$ & $1.048$ & $0.281$ \\ \hline
 $33$    &   $3.1$     &  $0.117$   &  $0.829$  & $0.713$ & $1.051$ & $0.531$ \\ \hline
 $34$    &   $2.2$     &  $0.200$   &  $0.764$  & $0.564$ & $1.065$ & $0.902$ \\ \hline
 $35$    &   $1.6$     &  $0.358$   &  $0.629$  & $0.271$ & $1.093$ & $1.595$ \\ \hline
 $35.2$  &   $1.53$    &  $0.426$   &  $0.565$  & $0.139$ & $1.100$ & $1.894$ \\ \hline
\end{tabular}
\caption{$T=1$, $n=1$ and $\delta=1/3600$, then $\phi_{min}=30^o$ and 
$\phi_{max}=35,3^o$. The lengths are in metres and the velocity in metres/second.}
\label{tabla2}
\end{center}
\end{table}

\begin{table}[h]
\begin{center}
\begin{tabular}{|c|c|c|c|c|c|c|} \hline
$\phi_0$ & $\tilde{E}$ & $H_{min}$  & $H_{max}$ &   $H$   & $L$     & $v_{\theta}$  \\ \hline \hline
 $14.4$  &   $61.7$    &  $0.0017$  &  $1.150$  & $1.148$ & $1.186$ & $0.007$ \\ \hline
 $14.8$  &   $9.2$     &  $0.029$   &  $1.142$  & $1.113$ & $1.186$ & $0.120$ \\ \hline
 $15.2$  &   $5$       &  $0.074$   &  $1.128$  & $1.054$ & $1.190$ & $0.299$ \\ \hline
 $15.6$  &   $3.4$     &  $0.134$   &  $1.102$  & $0.968$ & $1.202$ & $0.535$ \\ \hline
 $16$    &   $2.3$     &  $0.236$   &  $1.047$  & $0.811$ & $1.226$ & $0.921$ \\ \hline
 $16.5$  &   $1.66$    &  $0.462$   &  $0.885$  & $0.423$ & $1.275$ & $1.763$ \\ \hline
\end{tabular}
\caption{$T=1$, $n=2$ and $\delta=1/60$, then $\phi_{min}=14.35^o$ and 
$\phi_{max}=16.63^o$. The lengths are in metres and the velocity in metres/second.}
\label{tabla3}
\end{center}
\end{table}


\newpage
\begin{thebibliography}{99}

\bibitem{andronov} A.A. Andronov, A.A. Vitt and S.E. Khaikhin,
{\it Theory of Oscillators}, Dover Publications (1996).

\bibitem{lopezruiz} R. Lopez-Ruiz and A.F. Pacheco,
"Sliding on the inside of a conical surface", 
Eur. J. Phys. {\bf 23}, 579-589 (2002).
 
\bibitem{arnold} V.I. Arnold, {\it Mathematical Methods of Classical Mechanics},
Springer, Berlin (1978). 

\bibitem{goldstein} H. Goldstein, {\it Classical Mechanics},
Addison-Wesley, Reading (1980).

\bibitem{boltzmann} (a) L. Boltzmann, "Entgegnung auf die W\"armetheoretischen 
Betrachtungen des Hrn. E. Zermelo", Wied. Ann. {\bf 57}, 778-784 (1896); \\
(b) id., "Zu Hrn. Zermelos Abhandling '\"Uber die mechanische Erkl\"arung irreversibler
Vorg\"ange' ", Wied. Ann. {\bf 60}, 392-398 (1897); \\
(c) id. "\"Uber einen mechanischen Satz Poincar.e's", Wien. Ber. {\bf 106}, 12-20 (1897).

\bibitem{cohen} E.G.D. Cohen, "Boltzmann and Statistical Mechanics", lecture given
by at the International Meeting "Boltzmann's Legacy - 150 years after his Birth",
organized by the Accademia Nazionale dei Lincei, May-1994, in Rome, and published
in: {\it Atti dell'Accademia Nazionale dei Lincei} (1997).


\end{thebibliography}
\end{document}